\documentstyle[Buckow]{article}
\begin{document}
\def\titleline{
Low-Energy Effective Action
\newtitleline
in N=4 Super Yang-Mills Theory}

\def\authors{I.L. Buchbinder}

\def\addresses{
Department of Theoretical Physics,\\
Tomsk State Pedagogical University,\\
Tomsk 634041, Russia}

\def\abstracttext{
We consider $N=4$ supersymmetric Yang-Mills theory formulated in terms
of $N=2$ superfields in harmonic superspace. Using the background field
method we define manifestly gauge invariant and $N=2$ supersymmetric
effective action depending on $N=2$ strength superfields and develop a
general procedure for its calculation in one-loop approximation.
Explicit form for this effective action is found for the case of
$SU(2)$ gauge group broken down to $U(1)$.
}
%%%%%%%%%%%   %%%%%%%%%%%%%   %%%%%%%%%%%%%%%%%%
\large
\makefront

Maximally extended $N=4$ supersymmetric Yang-Mills (SYM) theory
attracts much attention due to remarkable properties on quantum level
and profound links with modern string/brane activity. It is natural to
expect that an effective action describing many aspects of quantum
theory, such as symmetry breaking, low- and high-energy behaviour,
anomalies and so on, also must possess the many remarkable features in
$N=4$ SYM. Unfortunately, the effective action of $N=4$ SYM was studied
unsufficiently so far and ones know little both about its general
structure and about its form in different approximations.

One of the main obstacles to investigate the effective action in $N=4$
SYM is absence of adequate quantum formulation such a theory. In our
opinion, an efficient technique of carring out the calculations of
effective action in $N=4$ SYM preserving $N=4$ supersymmetry manifestly
is still undeveloped. The best we have at present is treatment of
$N=4$ SYM as $N=2$ SYM coupled to a specific (hypermultiplet) matter
and use the methods of $N=2$ supersymmetric quantum field theory.

Recently Dine and Seiberg \cite{1} has shown that part of low-energy
effective action of $N=4$ SYM depending only on $N=2$ superfield
strengths $W$ and $\bar{W}$ can be found exactly up to numerical factor
on the base of restrictions imposed by $N=4$ supersymmetry. To be more
precise, it was shown that low-energy effective action
$\Gamma[W,\bar{W}]$ for $N=4$ $SU(2)$ theory is expressed in terms of
non-holomorphic effective potential ${\cal H}(W,\bar{W})$ as follows

\begin{equation}
\Gamma[W,\bar{W}]=\int d^4x  d^8\theta {\cal H}(W,\bar{W})
\label{1}
\end{equation}
where

\begin{equation}
{\cal H}(W,\bar{W})=c\,{\rm log}\frac{\bar{W}^2}{\Lambda^2}\,
{\rm log}\frac{W^2}{\Lambda^2}
\label{2}
\end{equation}
with some numerical coefficient $c$ and some scale $\Lambda$. The
effective action (\ref{1}) with ${\cal H}(W,\bar{W})$ (\ref{2}) is
$\Lambda$-independent.  The explicit calculation of the coefficient $c$
have been given in Refs \cite{7,8}. The generalizations for arbitrary
$SU(n)$ groups were considered in Refs \cite{9,10}. The present paper
is a brief overview of the approach to calculation of low-energy
effective action in $N=4$ SYM developed in Refs \cite{8,9}.

We consider $N=4$ SYM formulated in terms of $N=2$ superfields in
harmonic superspace \cite{2}. Such an approach is still the only one
operating with unconstrained $N=2$ superfields and explicitly
realizing $SU(2)_R$-symmetry of $N=2$ Poincare superalgebra. Manifestly
$N=2$ supersymmetric calculation of low-energy (holomorphic) effective
action in harmonic superspace was given in Refs \cite{3,4,6}.

From point
of view of $N=2$ supersymmetry, the $N=4$ SYM theory describes coupling
of $N=2$ vector multiplet to the hypermultiplet in the adjoint
representation. In harmonic superspace approach, the vector multiplet
is realized by an unconstrained analytic gauge superfield $V^{++}$
\cite{2} and hypermultiplet can be realized either by real
unconstrained analytic superfield $\omega$ ($\omega$-hypermultiplet) or
by a complex unconstrained analytic superfield $q^+$
($q$-hypermultiplet) \cite{2}. In the $\omega$-hypermultiplet
realization, the action of $N=4$ SYM theory reads

\begin{equation}
S[V^{++},\omega]=\frac{1}{2g^2}\int d^4x d^8\theta{\rm tr}W^2-
\frac{1}{2g^2}\int d\zeta^{(-4)}{\rm tr}\nabla^{++}\omega
\nabla^{++}\omega
\label{3}
\end{equation}
In the $q$-hypermultiplet representation, the $N=4$ SYM theory is given
by the action

\begin{equation}
S[V^{++},q^+,\breve{q}^+]=\frac{1}{2g^2}\int d^4x d^4\theta{\rm tr}W^2-
\frac{1}{2g^2}\int d\zeta^{(-4)}{\rm tr}q^{+i}\nabla^{++}q_i
\label{4}
\end{equation}
$$
q^+_i=(q^+,\breve{q}^+)\,,\quad q^{+i}=\varepsilon^{ij}q^+_j=
(\breve{q}^+,-q^+)
$$
Both models (\ref{3},\ref{4}) are manifestly $N=2$ supersymmetric.
However, they possess the two extra hidden supersymmetries \cite{2}
and, as a result, ones get $N=4$ supersymmetric theories with $N=4$ SYM
content. Of course, the theories (\ref{3},\ref{4}) are classically
equivalent. Our purpose is to describe a calculation of low-energy
effective action for the above theories.

The first step of calculation is background field quantization of the
model under consideration allowing to preserve manifest gauge
invariance and $N=2$ supersymmetry. Background field formulation for
$N=2$ supersymmetric field theories in harmonic superspace was
developed in Ref \cite{4} (see also \cite{5,6}). Within this
formulation, the one-loop effective action $\Gamma^{(1)}[W,\bar{W}]$
for the theories (\ref{3},\ref{4}) looks like

\begin{equation}
\Gamma^{(1)}[W,\bar{W}]=\frac{i}{2}
{\rm Tr}_{(2,2)}{\rm log}\stackrel{\frown}{\Box}-
\frac{i}{2}{\rm Tr}_{(4,0)}{\rm log}\stackrel{\frown}{\Box}
\label{5}
\end{equation}
where $\stackrel{\frown}{\Box}$ is the analytic d'Alambertian
introduced in Ref \cite{4} and the formal definition of the
${\rm Tr}_{(2,2)}{\rm log}\stackrel{\frown}{\Box}$ and
${\rm Tr}_{(4,0)}{\rm log}\stackrel{\frown}{\Box}$ are given in Ref
\cite{8}.

The second step is convenient path integral representation of the
effective action $\Gamma^{(1)}[W,\bar{W}]$ for special background. The
low-energy effective action (\ref{1}) depends only on $W$
$\bar{W}$, therefore it is sufficient to choose the background gauge
superfield on shell
\begin{equation}
{\cal D}^{\alpha(i}{\cal D}_\alpha^{\ j)}W=0
\label{6}
\end{equation}
where ${\cal D}_\alpha^{\ i}$ are the standard $N=2$ supercovariant
derivatives. In this case the effective action (\ref{5}) can
represented in the form
\begin{equation}
\exp(i\Gamma^{(1)}[W,\bar{W}])=\frac
{\int{\cal D}G^{++}\exp\left\{-\frac{i}{2}{\rm tr}\int d\zeta^{(-4)}
G^{++}\stackrel{\frown}{\Box}G^{++}\right\}}
{\int{\cal D}G^{++}\exp\left\{-\frac{i}{2}{\rm tr}\int d\zeta^{(-4)}
G^{++}G^{++}\right\}}
\label{7}
\end{equation}
where analytic integration superfield $G^{++}$ is constrained by
\begin{equation}
\nabla^{++}G^{++}=0
\label{8}
\end{equation}
The result (\ref{7}) is quite general and preserve manifestly all
symmetries of the theory.

The third step is transformation of the path integral (\ref{7}) to ones
over unconstrained $N=1$ superfields \cite{8}. This transformation can
be treated as some replacement of integration variables under the path
integral. We introduce also the $N=1$ projections of $W$: $\phi=W|$,
$2iW_\alpha={\cal D}_\alpha^{\ 2}W|$. The details of such a projection
are given in Ref \cite{8}. Here ${\cal D}_\alpha^{\ 2}\equiv{\cal
D}_\alpha^{\ i}\left.\right|_{i=2}$. As a result, ones obtain
\begin{eqnarray}
\exp(i\Gamma^{(1)})=\int{\cal D}\bar{V}{\cal D}V\exp\left\{
\frac{i}{2}\int d^8z{\rm tr}\bar{V}\Delta V\right\}\nonumber\\
\label{9} \\
\Delta={\cal D}^a{\cal D}_a-eW^\alpha{\cal D}_\alpha+
e\bar{W}_{\dot{\alpha}}\bar{{\cal D}}^{\dot{\alpha}}+e^2|\phi|^2
\nonumber
\end{eqnarray}
Here ${\cal D}_a$, ${\cal D}_\alpha$, $\bar{{\cal D}}_{\dot{\alpha}}$
are the $N=1$ supercovariant derivatives and $\Gamma^{(1)}$ depends on
$\phi$, $W_\alpha$, $\bar{\phi}$, $\bar{W}_{\dot{\alpha}}$, $V$ and
$\bar{V}$ are unconstrained $N=1$ complex scalar superfields.

The final step is calculation of $\Gamma^{(1)}$ (\ref{9}) in low-energy
limit. We consider the background gauge superfield corresponding to the
unbroken $U(1)$ subgroup of $SU(2)$ group in the Coulomb branch. In
this case the effective action $\Gamma^{(1)}$ depending on
$\phi$, $\bar{\phi}$, $W_\alpha$, $\bar{W}_{\dot{\alpha}}$ can be
written as follows
\begin{equation}
\Gamma^{(1)}=\int d^8zW^\alpha W_\alpha \bar{W}_{\dot{\alpha}}
\bar{W}^{\dot{\alpha}}
\frac{\partial^4 {\cal H}(\phi,\bar{\phi})}
{\partial\phi^2\partial\bar{\phi}^2}+\dots
\end{equation}
To calculate $\partial^4 {\cal H}(\phi,\bar{\phi})/
\partial\phi^2\partial\bar{\phi}^2$ we use $N=1$ superfield proper-time
technique introducing Schwinger kernel for the operator $\Delta$
(\ref{9}) (see the details of superfield proper-time technique in Ref
\cite{11}).
\begin{equation}
\Gamma^{(1)}=-i\int\limits_0^\infty\frac{ds}{s}e^{-i(e^2|\phi|^2-i
\varepsilon)s}\int d^8z {\cal U}(z,z|s)
\end{equation}
where for constant $W$ and $\bar{W}$ ones obtain
\begin{equation}
{\cal U}(z,z'|s)=\frac{i}{(4\pi is)^2}e^{is(eW^\alpha D_\alpha-
e\bar{W}_{\dot{\alpha}}\bar{D}^{\dot{\alpha}})}
\delta^4(\theta-\theta')e^{\frac{(x-x')^2}{4is}}
\end{equation}

It leads to
\begin{equation}
\partial^4 {\cal H}(\phi,\bar{\phi})/
\partial\phi^2\partial\bar{\phi}^2=(4\pi\bar{\phi}\phi)^{-2}
\end{equation}

One can
easily find a general solution of this equation and restore the
function ${\cal H}(W,\bar{W})$. We finally get
\begin{equation}
{\cal H}(W,\bar{W})=\frac{1}{4(4\pi)^2}\,
{\rm log}\frac{W^2}{\Lambda^2}\,
{\rm log}\frac{\bar{W}^2}{\Lambda^2}\,
\end{equation}
Thus the coefficient $c$ in (\ref{2}) is equal to $1/(4(4\pi)^2)$. All
details of the calculations can be found in Ref \cite{8}.

To conclude, we have formulated a general procedure of calculating the
low-energy effective action depending on $N=2$ strength superfields
for $N=4$ SYM theories.

\vskip0.5cm
\noindent
{\large \bf Acknowledgements}

\smallskip
\noindent
I am grateful to S.M. Kuzenko for fruitful collaboration and to
E.I. Buchbinder, E.A. Ivanov and B.A. Ovrut for numerous discussions.
The work was supported in part by RFBR grant, INTAS grant, project
$N^o$ 96-0308, RFBR-DFG grant, project $N^o$ 96-02-00180,
GRACENAS grant, project $N^o$ 97-6.2-34. It is pleasure to thank the
Organizing Commettee of the Buckow Symposium for the warm hospitality.

\end{document}